\documentclass[twocolumn,showpacs,preprintnumbers,amsmath,amssymb,superscriptaddress]{revtex4-1}

\usepackage{graphicx}
\usepackage{dcolumn}
\usepackage{bm}
\usepackage{color}

\begin{document}


\title{Coefficient of restitution for wet particles}

\author{Frank Gollwitzer}
\author{Ingo Rehberg}
\affiliation{Experimentalphysik V, Universit\"at Bayreuth, 95440 Bayreuth, Germany}
\author{Christof A. Kruelle}
\affiliation{Maschinenbau und Mechatronik, Hochschule Karlsruhe - Technik und Wirtschaft, D-76133 Karlsruhe, Germany}
\author{Kai Huang}
\email{kai.huang@uni-bayreuth.de}
\affiliation{Experimentalphysik V, Universit\"at Bayreuth, 95440 Bayreuth, Germany}

\date{\today}

\begin{abstract}

The influence of a wetting liquid on the coefficient of restitution (COR) is investigated experimentally by tracing freely falling particles bouncing on a wet surface. The dependence of the COR on the impact velocity and various properties of the particle and the wetting liquid is presented and discussed in terms of dimensionless numbers that characterize the interplay between inertial, viscous, and surface forces. In the Reynolds number regime where the lubrication theory does not apply, the ratio of the film thickness to the particle size is found to be a crucial parameter determining the COR.

\end{abstract}


\maketitle

\section{Introduction}

The coefficient of restitution (COR), first introduced by Newton \cite{Newton1687} as the ratio between the relative rebound and impact velocities of a binary impact, has been a subject of continuous interest over centuries, along with the development of elastic \cite{Hertz1882, Love1927}, viscoelastic \cite{Rami99} and plastic theories \cite{Tabor1948, Johnson85}. It characterizes the energy dissipation associated with the impact, which plays a key role in understanding the collective behavior of macroscopic particles, i.e.\ the dynamics of granular matter \cite{Nagel96, Duran00}. This is largely due to the fact that the dissipative nature of granular matter arises from the inelastic collisions at the particle level.

Due to its omnipresence in nature and various industries, granular matter has drawn great attention from both physical and engineering communities in the past decades \cite{pg09}. Concerning the modeling of granular matter, an appropriate collision model is essential for the successful implementation of kinetic or hydrodynamic theories to granular matter \cite{Bril96, bril04, Jenkins83, Goldhirsch03}, see for example the dynamics of Saturn's rings \cite{Spahn06}, or the pattern formation under vertical agitation \cite{Bizon98}. Despite those successful examples for dry granular matter, a continuum description for wet granular matter, which considers the cohesion arising from the wetting liquid phase, is still far from established  \cite{Herminghaus05, Pouliquen06}. Therefore, in order to provide a solid basis for a continuum modeling of wet granular flow -- for example to describe natural disasters such as debris flow -- a thorough understanding of the dynamics associated with wet impacts is desirable.

\begin{figure}[b]
\includegraphics[width = 0.4\textwidth]{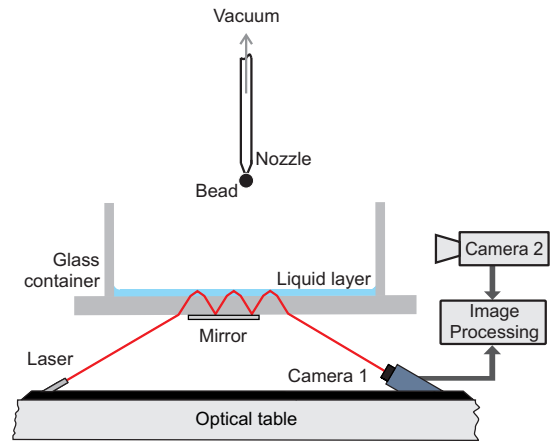}
\caption{\label{setup}(color online) Sketch of the experimental setup. The bouncing of the glass bead, initially held by the vacuum nozzle, on the glass container is recorded with a high speed camera (Camera 2). The thickness of the liquid layer is monitored by detecting the laser beam (red line) reflected from the liquid surface with Camera 1.}
\end{figure}

\begin{figure*}[t]
\includegraphics[width =0.8\textwidth]{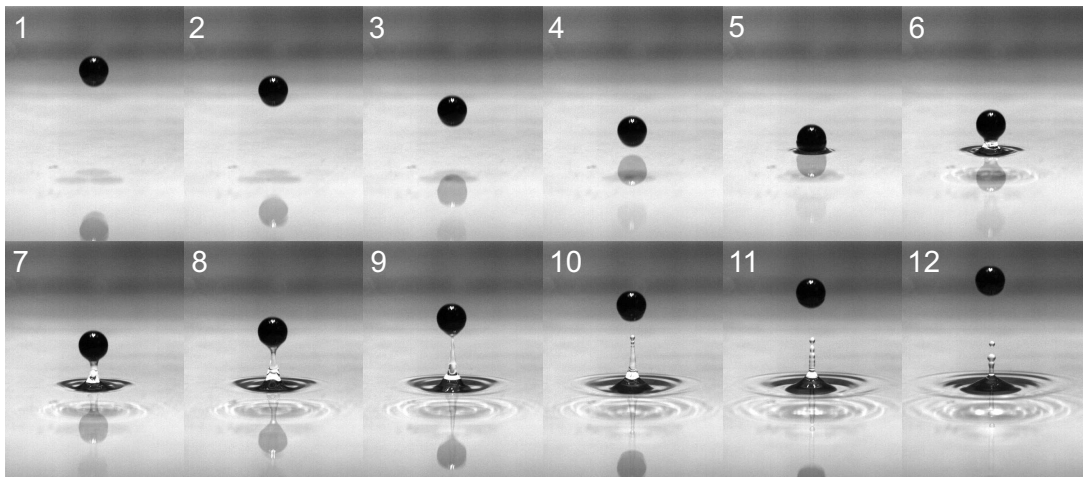}
\caption{\label{snapshot} A series of snapshots captured with a frame rate of $450$\,Hz showing a $4$\,mm glass bead bouncing on a glass plate covered with a $1$\,mm water film.}
\end{figure*}

With the development of pharmaceutics, mining and food industries, the COR for wet impacts has become an important issue for the engineering community in terms of decoding the underlying physics associated with the agglomeration of particles with liquid binders. The pioneering work by Rumpf \cite{Rumpf62} half a century ago has included a detailed description of the capillary force of a pendular bridge and treated it as the dominating cohesive force in determining the continuum properties of wet granular matter, e.g. the tensile strenth \cite{Pietsch68}. Later on, the viscous force has been found to play an important role in typical granulation processes, too \cite{Ennis90, Ennis91, Iveson96, Iveson98, Iveson01, Fu04, Antonyuk09, Mueller11}. And a dynamic liquid bridge could be an order of magnitude stronger than a quasi-static one \cite{Mazzone87, Murase04}. Binary as well as three body impacts of particles with viscous liquid coating have been extensively investigated by experiments and models using lubrication theory \cite{Davis86, Davis88, Thornton98, Davis02, Kantak06, Donahue10, Donahue10b}.


Despite all those investigations, a well tested collision law suitable for modeling the dynamics of wet granular behavior \cite{Ulrich09, Fingerle08, Fingerle08b}, as well as a comprehensive knowledge of the energy dissipation associated with the impact, is still lacking. In the current work, the COR of a ball bouncing back from a flat lubricated surface is investigated as a function of the impact velocity, various particle sizes and liquid properties. From this, the kinetic energy dissipated during the impact process is derived and discussed within the framework of existing models.

\section{Experimental setup and procedure}

Figure~\ref{setup} shows a sketch of the experimental setup used for the COR measurements. Spherical glass beads (SiLiBeads type P) with a diameter range from $D=2.8$\,mm to $10$\,mm, roughness $\approx5$\,$\mu$m, and density $\rho_{g}=2.58$\,${\rm g}/{\rm cm}^3$ are used in the experiments. By controlling the pressure in the vacuum nozzle, we allow an initially wet particle to fall freely onto a wet glass container ($20$\,cm$\times$\,$5$\,cm). The initial falling height is adjusted from $20$\,mm to $145$\,mm, corresponding to an initial impact velocity range from $\approx0.3$\,m/s to $\approx1.7$\,m/s. Three types of wetting liquids with various properties, as shown in Table\,\ref{tab:liq_properties}, are used. The bottom of the container is thick enough ($2$\,cm) to avoid any influence on the COR \cite{Sondergaard90} for the range of particle size used. It is leveled within $0.03$\,degrees, so that bouncing on various positions in the container explores a similar liquid layer thickness. 

The layer thickness $\delta$ used in the current investigation ranges from $75\,\mu$m to $1$\,mm. It is measured by detecting the shift of a laser beam reflected from the surface of the liquid and the glass plate with a CCD camera (Camera 1, Lumenera Lu135). The mirror attached to the bottom of the container creates multiple reflections of the laser beam, in order to enhance the sensitivity of the device. The length of the mirror ($7.8$\,cm) is chosen as a compromise between the sensitivity and the field of view. By fixing the container, laser and the camera on a leveled optical table, the error of the film thickness measurement could be minimized to a satisfactory level ($<10$\,$\mu$m).

\begin{table}
\caption{\label{tab:liq_properties}%
Material properties of the wetting liquids at $20\,^{\circ}{\rm C}$. M5 and M50 correspond to two types of silicone oil from Carl Roth.}
\begin{ruledtabular}
\begin{tabular}{cccc}
\textrm{}&
\textrm{Density}&
\textrm{Viscosity}&
\textrm{Surface tension}\\
\textrm{}&
\textrm{(kg/m$^3$)}&
\textrm{(mPa\,s)}&
\textrm{(mN/m)}\\
\colrule
Water & 998 & 1.0 & 72.8\\
M5 & 925 & 4.6 & 19.2\\
M50 & 965 & 48 & 20.8\\
\end{tabular}
\end{ruledtabular}
\end{table}

\begin{figure}
\includegraphics[width = 0.4\textwidth]{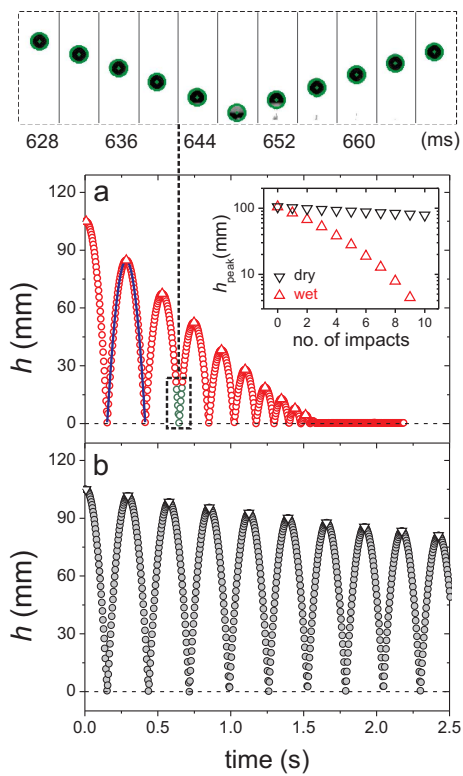}
\caption{\label{trj}(color online) Trajectories of a particle bouncing on a wet (a) and a dry (b) surface after image processing. The image sequence in the upper panel represents a fraction of the wet trajectory (a) with superimposed centers and boundaries of the sphere. The blue line in (a) corresponds to a parabolic fit to the trajectory after the first rebound. The peak positions of the trajectories $h_{\rm peak}$ obtained from the fits are marked with triangles in (a) and upside down triangles in (b). The inset in (a) shows $h_{\rm peak}$ as a function of the number of impacts.}
\end{figure}

To obtain the impact and rebound velocities, the bouncing of the particle is recorded by a fast camera (Photron Fastcam Super 10K) and subsequently applied to an image processing procedure. A close view of the colliding event, as shown in Fig.\,\ref{snapshot}, clearly demonstrates the important role that the wetting liquid plays during the impact. As the sphere hits the liquid surface, a circular wave front occasionally accompanied with a splash will be generated. As the ball rebounds from the surface, a liquid bridge will form between the sphere and the liquid surface, which continuously deforms and elongates until it ruptures at a distance larger than the particle diameter. Associated with the rupture event, satellite droplets may form, which bounce on the liquid surface and coalesce partially into smaller droplets \cite{Couder05b,Blanchette06}. Obviously, the formation of wave fronts, deformation and rupture of liquid bridges, the viscous force, and the added mass to the sphere due to the wetting liquid will all contribute to the mechanical energy reduction of the impacting particle, which in turn leads to a smaller COR compared with dry impacts.

Figure~\ref{trj} illustrates the influence of wetting by providing a comparison between the trajectories obtained from wet and dry impacts. The particle diameter is $D=5.5\,$mm, and the film thickness of the silicone oil M5 is $\delta=225\,\mu$m in the wet case. To determine the location of the sphere centers, the image processing procedure employs a Hough transformation \cite{Kimme75} (upper panel of Fig.\,\ref{trj}). Subsequently, each bouncing trajectory is extracted and subjected to a parabolic fit (see the solid line in Fig.\,\ref{trj}(a) as an example), in order to obtain the peak position $h_{\rm peak}$ and the impact velocity.

If the normal COR, also represented as $e_{\rm n}$, is independent of the impact velocity, the velocity after the $i$th rebound will be related to the first impact velocity $v_{\rm 0}$ by $v_{\rm i}=e_n^i v_{\rm 0}$. This leads to a linear decay of the peak height $h_{\rm peak}$ with the number of impacts $i$ in semi-log plot, according to $\log_{10} h_{\rm peak}=\log_{10} h_{0}+2i\log_{10} e_n$, with $h_{\rm peak}\propto v_i^2$. The initial falling height $h_0$ and $e_n$ determine the offset and slope of this line. As shown in the inset of Fig.\,\ref{trj}(a), the logarithm of $h_{\rm peak}$ decreases linearly with the number of impacts for dry impacts, indicating that the normal COR stays almost constant for the number of impacts measured here. In a recent work on dry impacts \cite{Montaine11}, a more detailed analysis reveals that the dry COR decreases slightly with the increase of $v_{\rm impact}$. However, this dependence is much weaker than the one for wet impacts, on which we are focusing here. In this case, the variation of the slope indicates that the COR for wet impacts decreases strongly with the number of impacts, i.e.\ with the impact velocity.

Even though using $h_{\rm peak}$ gives a practical analysis of the COR, this method may suffer the influence from interstitial air. Therefore the normal COR is obtained, based on its definition, from the ratio between the fitted rebound and impact velocities for the rest of the paper.

\section{Experimental results}

\begin{figure*}[t]
\includegraphics[width = 0.4\textwidth]{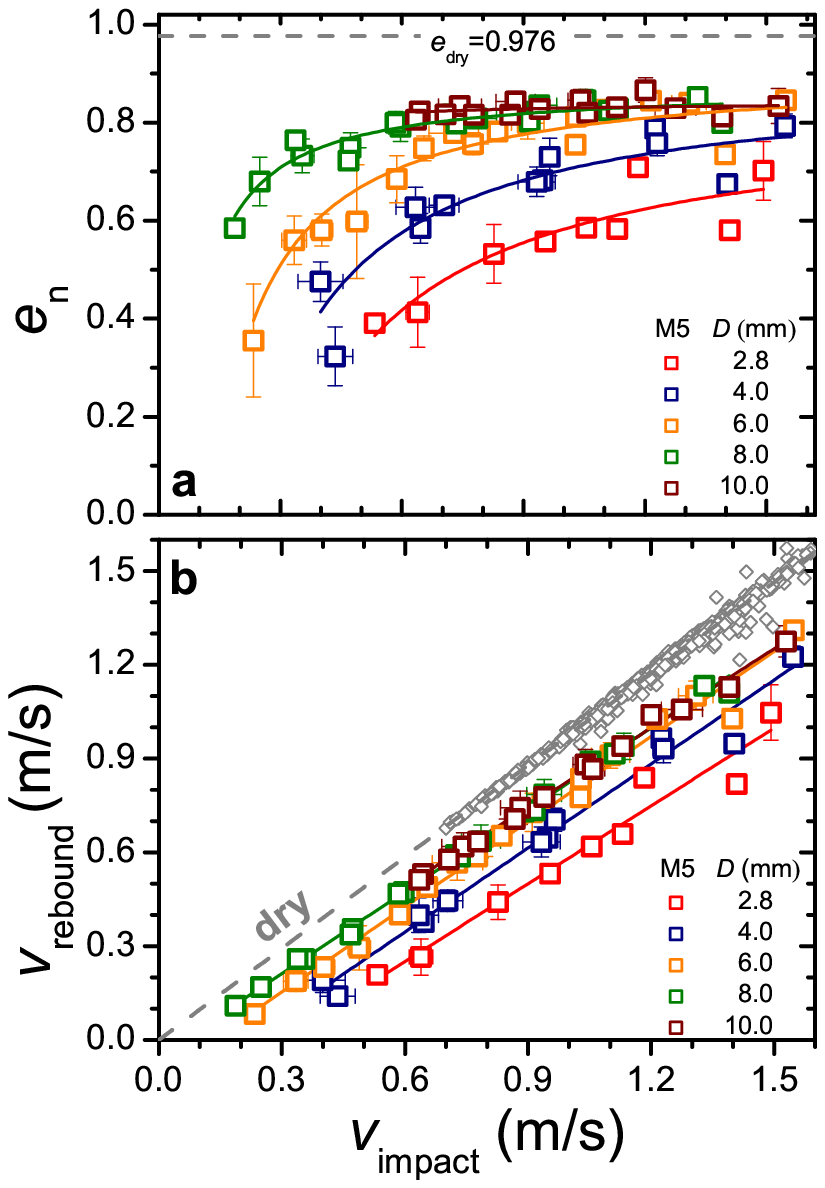}
\includegraphics[width = 0.4\textwidth]{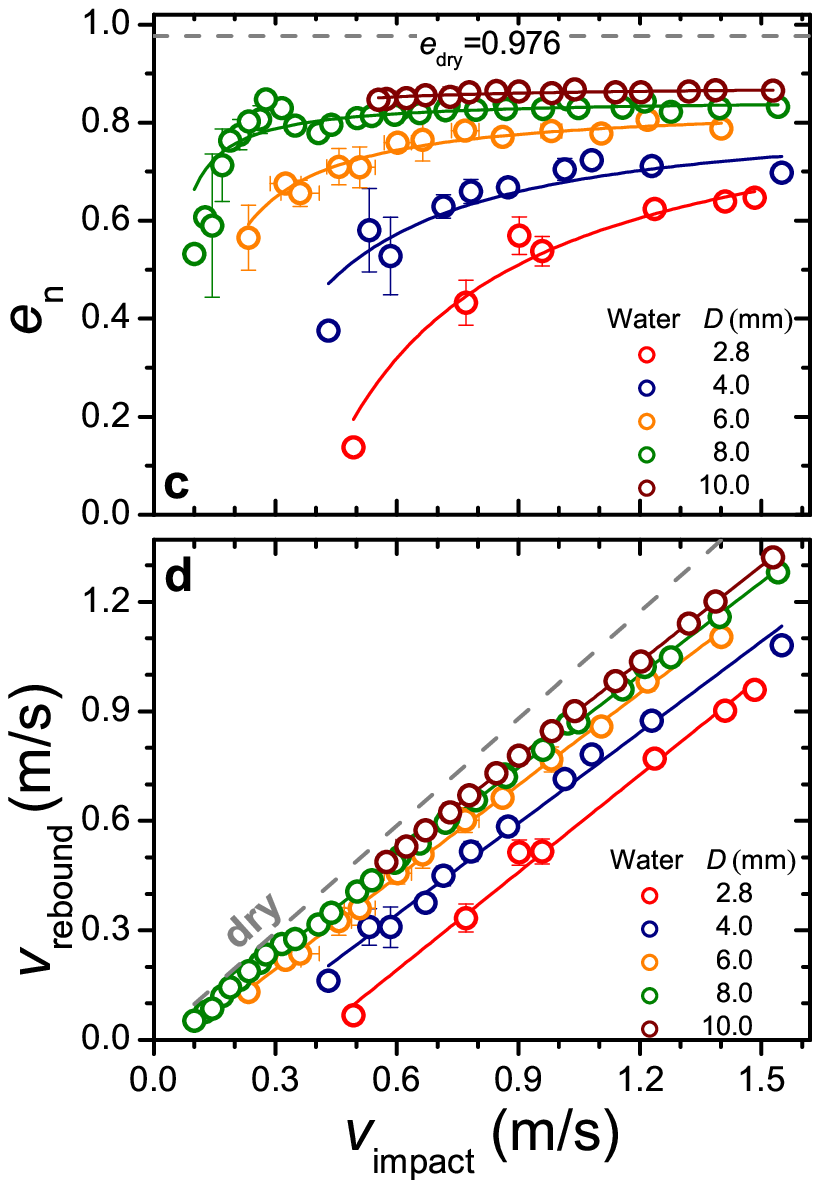}
\caption{\label{vivo2}(color online) Normal restitution coefficient $e_{\rm n}$ and rebound velocity $v_{\rm rebound}$ as a function of impact velocity $v_{\rm impact}$ for the impacts of particles with various diameters $D$ on silicone oil (left column) and water (right column) wetting surfaces with fixed film thickness $\delta = 1$\,mm. The solid lines in the lower panels are linear fits to the data and their representatives are shown in the upper panels as a guide to the eyes. The dashed gray lines in the upper panels represent the normal restitution coefficient $e_{\rm dry}=0.976\pm0.001$ for dry impacts, which is obtained by a linear fit of the data for all particle sizes (gray diamonds shown in b). Error bars smaller than the symbol size are not shown.}
\end{figure*}

\begin{figure}
\includegraphics[width = 0.4\textwidth]{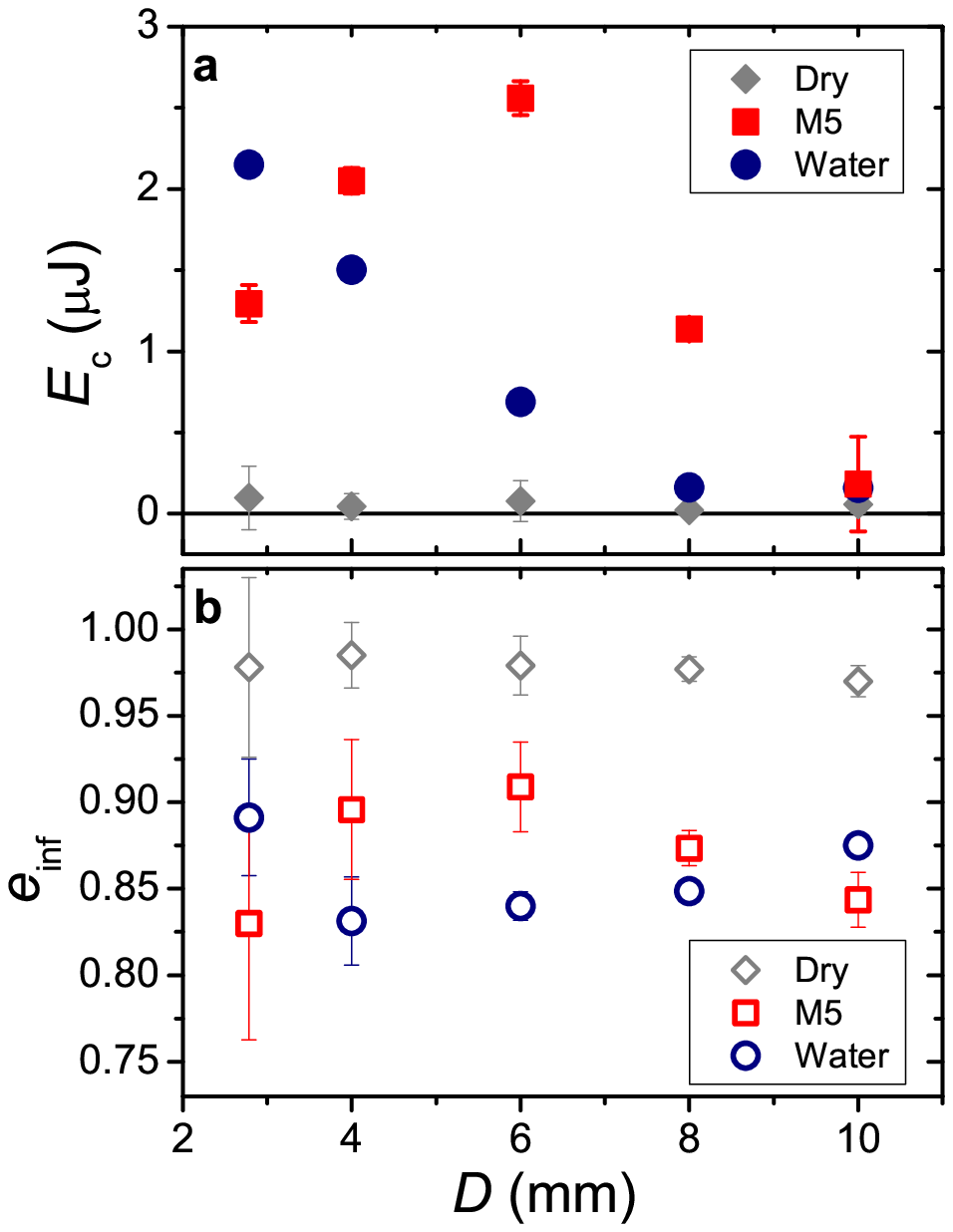}
\caption{\label{LinFit1} The critical energy $E_{\rm c}$ and the saturated value of the COR $e_{\rm inf}$, obtained from linear fits in Fig.\,\ref{vivo2}(b), as a function of particle diameter $D$. The solid line at $E_{\rm c}=0$ is a guide to the eyes.}
\end{figure}

Figure\,\ref{vivo2} shows the dependence of the COR on the impact velocity $v_{\rm impact}$ and various particle diameters for both silicone oil (M5) and water wetting. Qualitatively, the same trend for the impact velocity dependence is observed: The COR grows initially with $v_{\rm impact}$ and saturates at a certain value, as shown by the guided lines in the upper panel. In the lower panel of Fig.~\ref{vivo2}, the rebound velocity $v_{\rm rebound}$ is plotted as a function the impact velocity $v_{\rm impact}$. Similar to the case without wetting liquid (shown as a gray dashed line), $v_{\rm rebound}$ grows linearly with $v_{\rm impact}$ for all parameters used here. Different from the dry impacts, the fitted line has an offset with the x-axis, which explains the growth of the COR with $v_{\rm impact}$. Fitting the data with $v_{\rm rebound}=e_{\rm inf}(v_{\rm impact}-v_{\rm c})$ gives rise to two parameters that characterize the impact velocity dependence: A slope $e_{\rm inf}$ corresponding to the COR at infinite $v_{\rm impact}$, i.e.\ the saturated value of the COR, and an offset $v_{\rm c}$ corresponding to a critical energy $E_{\rm c}$ below which no rebound would occur. $E_{\rm c}=mv_{\rm c}^2/2$ is obtained from the intersection $v_{\rm c}$ of the linear fits shown in the lower panels of Fig.\,\ref{vivo2} with the x-axis, where $m$ is the mass of the particle. 

Besides the impact velocity, the COR is also found to be dependent on the size of the particles. For fixed $v_{\rm impact}$, the COR decreases systematically with particle diameter for both silicone oil and water wetting. Since the COR is related to the fraction of kinetic energy retained after the impact, the growth of the COR with $D$ indicates that the energy dissipation from the wetting liquid grows slower with $D$ than the inertia ($\propto D^3$) of the particles. 

Figure~\ref{LinFit1} shows the dependence of the parameters $e_{\rm inf}$ and $E_{\rm c}$ from the linear fits on the particle diameter for both wetting liquids. For the dry impacts, $E_{\rm c}$ stays constantly at 0 within the error bar. In contrast, the critical energy for wet impacts is on the order of few $\mu$J. It shows a monotonic decay for water wetting, and a more complicated relationship for the case of silicone oil M5 wetting. As shown in Fig.\,\ref{LinFit1} (b), $e_{\rm inf}$ -- the upper limit of the COR -- varies from $0.8$ to $0.9$, and is generally smaller than $e_{\rm dry}$. This indicates that the ratio between the energy dissipation from the wetting liquid, $\Delta E_{\rm wet}$, and the kinetic energy at impact $E_{\rm i}$ will not diminish as $v_{\rm impact}$ grows. For both silicone oil (M5) and water wetting, $e_{\rm inf}$ shows similar values with weak dependence on the particle sizes, although M5 silicone oil is 5 times more viscous than water. For dry impacts, the slope $e_{\rm inf}$ shows a weak dependence on the particle size. Linear fitting over the data from various $D$ suggests an averaged $e_{\rm dry}=0.976$, as shown in Fig.\,\ref{vivo2}. For wet impacts, the error bar for $e_{\rm inf}$ is larger as $D$ decreases. This is presumably due to the larger influence from the liquid film, which may lead to a larger inertial effect from the liquid flow and a more complex energy dissipation scenario. Thus we keep the liquid film thickness within 1\,mm for the COR dependence on the liquid properties shown below.

\begin{figure}
\includegraphics[width = 0.45\textwidth]{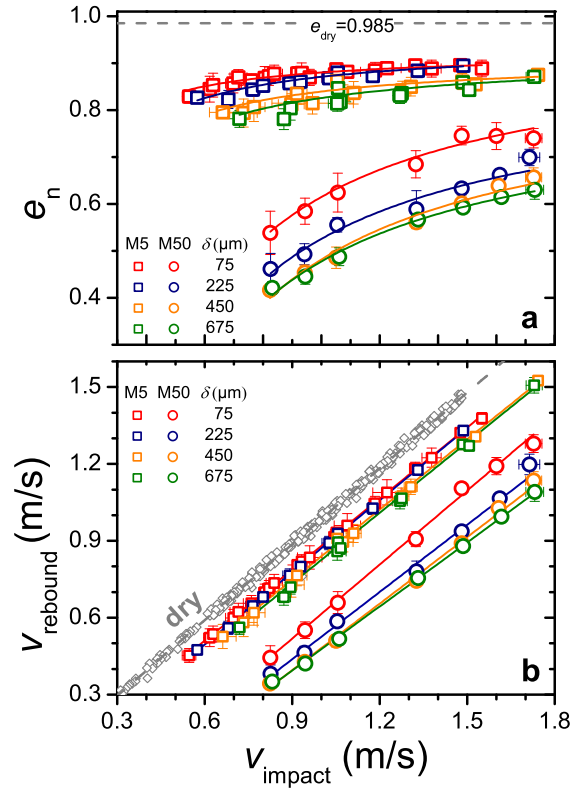}
\caption{\label{vivo}(color online) The normal restitution coefficient $e_n$ (a) and rebound velocity $v_{\rm rebound}$ (b) as a function of impact velocity $v_{\rm impact}$ for impacts of a glass bead with $D=5.5\,$mm on dry and wet surfaces covered with silicone oil M5 and M50. $\delta$ denotes the film thickness. The error bars correspond to the statistical error over 10 runs of experiments for the wet impacts. Solid lines in (b) are linear fits to the corresponding data. Their representatives are shown in (a) as a guide to the eye. Squares and dots correspond to M5 and M50, respectively. For dry impacts, the restitution coefficient $e_{\rm dry}$ is 0.985 with an error of 0.001.}
\end{figure}

\begin{figure}
\includegraphics[width = 0.4\textwidth]{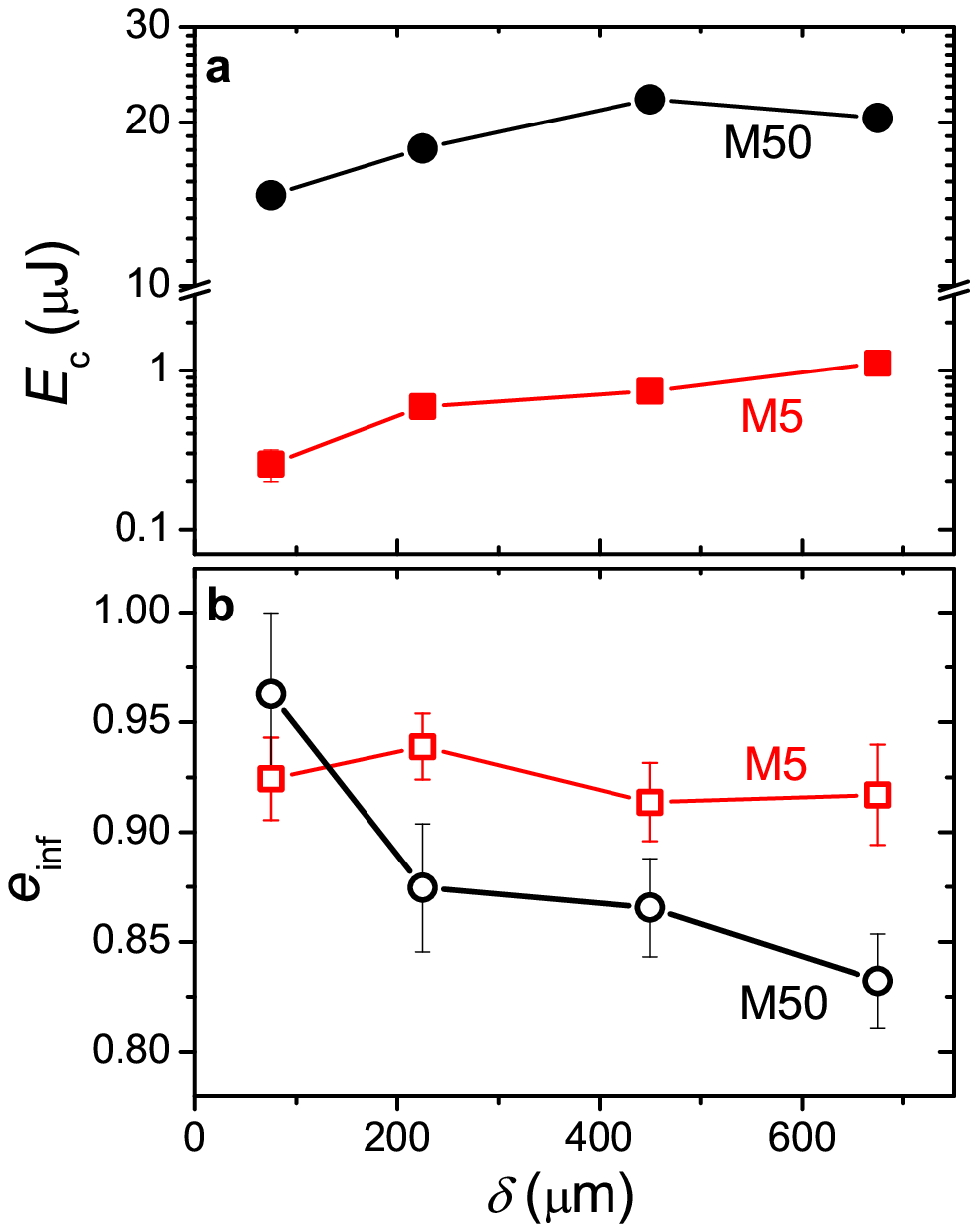}
\caption{\label{LinFit2} The critical energy $E_{\rm c}$ and the saturated value of the COR $e_{\rm inf}$ as a function of the film thickness $\delta$. $E_{\rm c}$ is obtained from the intersection $v_{\rm c}$ of the linear fits shown in the lower panel of Fig.\,\ref{vivo} with the x-axis. $e_{\rm inf}$ corresponds to the slope of these fits.}
\end{figure}

The influence of the liquid film thickness $\delta$ and the dynamic viscosity $\eta$ on the wet impacts is presented in Fig.\,\ref{vivo}. Here only silicone oil with various viscosities is chosen because of two reasons. Firstly, it wets the glass surface better than water due to its low surface tension and contact angle, and thus facilitates investigations on relatively thin liquid film. Secondly, the two types of silicone oil with various viscosities have a similar surface tension and density, which facilitates comparisons. Each data point shown here corresponds to an average of 10 runs of experiments with various initial falling heights and the error bar represets the statistical error.

Similar to the results shown in Fig.\,\ref{vivo2}, the rebound velocity increases linearly with the impact velocity with an offset with the x-axis (as shown in Fig.\,\ref{vivo}(b)), leading to a growth of the COR with $v_{\rm impact}$ towards a saturated value $e_{\rm inf}$ smaller than $e_{\rm dry}=0.985$. Note that $e_{\rm dry}$ obtained here for the 5.5\,mm particle is slightly larger than the one in Fig.\,\ref{vivo2}, which presumably arises from the variation of the COR on particle diameter shown in Fig.\,\ref{LinFit1}(b). A comparison between both wetting liquids shows that the $v_{\rm impact}$ dependence of the COR is more prominent for more viscous wetting liquid M50, as the larger offset from the linear fits indicates. As the film thickness $\delta$ increases, $e_{\rm n}$ decreases systematically for both wetting liquids, because the viscous damping force is effective over a larger distance. Further tests with increased film thickness up to 1.35\,mm yield qualitatively the same $v_{\rm impact}$ dependence.



As shown in Fig.\,\ref{vivo}(b), the relation between $v_{\rm rebound}$ and $v_{\rm impact}$ also represents the influence from the thickness and viscosity of the wetting liquid. For wet impacts, $v_{\rm rebound}$ decreases systematically with the liquid film thickness $\delta$ at a certain $v_{\rm impact}$. As the liquid viscosity increases by an order of magnitude (from M5 to M50), this trend is more prominent, indicating the crucial role played by the viscous damping. Fitting the growth of $v_{\rm rebound}$ with $v_{\rm impact}$ with a straight line again gives rise to two parameters: A slope $e_{\rm inf}$ that is smaller than $e_{\rm dry}$ and a threshold energy $E_{\rm c}$ below which no rebound would occur. As shown in Fig.~\ref{LinFit2}(a), this threshold is, for M50, more than an order of magnitude larger than that for M5. This suggests the dependence of $E_{\rm c}$ on the viscosity. As shown in Fig.~\ref{LinFit2}(b), the slope $e_{\rm inf}$ is not strongly influenced by viscosity compared with $E_{\rm c}$. For relatively thin film wetting, $e_{\rm inf}$ could be the same within the error bars. The slope $e_{\rm inf}$ stays constant within the range of film thickness and decays slightly for more viscous silicone oil M50 wetting.



\section{Scaling with the Stokes number}

The above experimental results indicate that the COR depends strongly on the impact velocity, particle sizes, and various liquid properties. In order to explore the relation between the COR and all these parameters, it is essential to have a proper classification of the parameters in terms of dimensionless quantities that characterize the relation between inertia, viscous and capillary effects. In the case where the viscous force dominates, the lubrication theory has been applied to explain the dynamics of wet impacts \cite{Davis86, Ennis91, Davis02, Donahue10b}. In such a case, the Stokes number is used to characterize the dependence of the COR on various control parameters. The Stokes number ${\rm St}=\rho_{\rm g}Dv_{\rm impact}/9\eta$ is defined as the ratio between the inertia of the particle and the viscosity of the wetting liquid, where $\rho_{\rm g}$ is the density of the glass beads. Normally, this case is justified by the criterion ${\rm Re}\ll1$ \cite{Davis02}. The Reynolds number ${\rm Re}$ is defined as $\rho_{\rm l}\delta v_{\rm impact}/\eta$, where $\rho_{\rm l}$ and $\delta$ are the density and the thickness of the wetting liquid correspondingly. This implies that either the liquid is highly viscous, or the film thickness is small. Within this limit, the contribution from the wetting liquid to the total energy dissipation is mainly due to viscous damping. Although the range of Reynolds numbers for the current investigation (up to $\approx 10^3$) suggests that the role that the viscous force plays may not always be prominent, we still use the Stokes number to rescale the dependence of the COR on various parameters as a starting point.

\begin{figure}
\includegraphics[width = 0.4\textwidth]{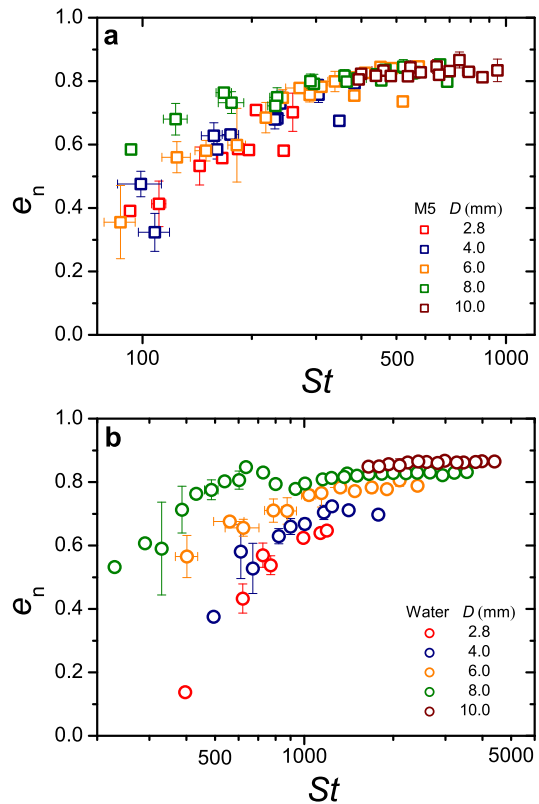}
\caption{\label{St1}(color online) Normal restitution coefficient $e_{\rm n}$ as a function of the Stokes number ${\rm St}$ for both silicone oil (a) and water (b) films. Parameters are the same as in Fig.~\ref{vivo2}.}
\end{figure}


Figure\,\ref{St1} is a re-plot of the data in Fig.\,\ref{vivo2}(a) and (c) in the $e_{\rm n}$ -- St plane. For wetting liquid silicone oil M5, which has a kinematic viscosity $5$ times that of water, the rescaling with the Stokes number yields better overlapping than that for water wetting. This could be attributed to the lower Re range (${\rm Re}=20$\,--\,$360$) for the case of silicone oil M5 wetting, which leads to more prominent influence from the viscosity. As shown in Fig.\,\ref{St1}(a), data for various $D$ show a general trend of initial growth from ${\rm St}\approx 100$ to $500$, followed by a saturation to $e_{\rm inf}$ between $0.8$ and $0.85$. Concerning the case of water wetting (corresponding to ${\rm Re}=100$\,--\,$1800$), the scatter of the data obtained with various particle sizes (shown in Fig.\,\ref{St1}(b)) is much more prominent than for the case of silicone oil M5 wetting. Although the trend of a significant growth followed by a saturated value persists, both the slope of increase and the saturated value differ as $D$ varies.

From another point of view, Fig.\,\ref{St1} also reveals a relatively small difference of the COR between M5 and water wetting, even though the corresponding viscosity ratio is $5$. This result indicates that the COR is also determined by other liquid properties. As an example, the surface tension of water is much larger than that of M5, which may lead to a larger energy dissipation from the formation of capillary waves and the break of capillary bridges upon rebound. In order to study the influence from viscosity, we focus on the results from silicone oil M5 and M50 (shown in Fig.\,\ref{vivo}), which have similar surface tension and density (see Table\,\ref{tab:liq_properties}), in the following part of the section.

\begin{figure}
\includegraphics[width = 0.45\textwidth]{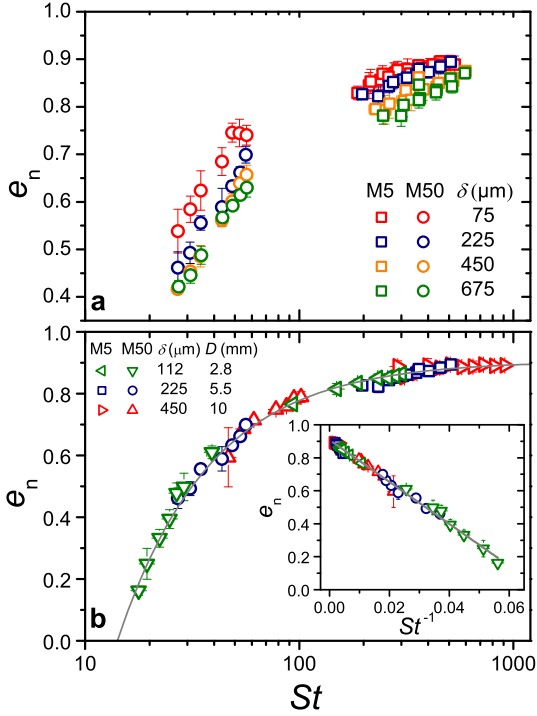}
\caption{\label{St}(color online) The normal restitution coefficient as a function of the Stokes number ${\rm St}$. (a) corresponds to the data shown in Fig.\ref{vivo}. (b) corresponds to the data with the dimensionless film thickness $\delta/D$ roughly constant. The other parameters are the same as in Fig.\,\ref{vivo}. The error bars correspond to the statistical error of 10 runs of experiments. The solid curve in (b) corresponds to the master curve $e_{\rm n}=0.908(1-14.00/{\rm St})$, which is obtained from a linear fit to all the data shown in the inset.}
\end{figure}

Figure\,\ref{St}(a) shows the COR as a function of the St number for the data shown in Fig.\,\ref{vivo}(a). The COR grows dramatically at small ${\rm St}$, which corresponds to the data of the more viscous silicone oil M50 wetting case, and saturates at larger ${\rm St}$. For various film thickness $\delta$, this trend is qualitatively the same. This trend, as well as the data scattering at low St, is also comparable to the results with various particle diameter shown in Fig.\,\ref{St1}(a). Quantitatively, the saturated value $e_{\rm inf}$ decreases as the film thickness $\delta$ grows, suggesting further dimensionless parameters associated with $\delta$ have to be considered.

This parameter is chosen as the dimensionless length scale $\tilde{\delta}=\delta/D$, because it ties the Stokes number with the Reynolds number of the wetting liquid. According to this definition, the ratio between the Reynolds number and the Stokes number is ${\rm Re}/{\rm St}=9\tilde{\delta}\tilde{\rho}$, where $\tilde{\rho}=\rho_{\rm l}/\rho_{\rm g}$ is the density ratio between the liquid and the particle.

In Fig.\,\ref{St}(b), further experiments with the restriction $\tilde{\delta}\approx 0.04$ are presented. In contrast to Fig.\,\ref{St}(a), the data from various film thicknesses coincide over a wide range of ${\rm St}$ if $\tilde{\delta}$ is fixed. It also gives rise to a master curve $e_{\rm n}=e_{\rm inf}(1-St_{\rm c}/{\rm St})$, as indicated clearly in the inset. The linear fit yields $e_{\rm inf}=0.908\pm0.002$, and a critical Stokes number ${\rm St_c}=14.00\pm0.20$. Therefore, the usage of the Stokes number as a control parameter could be extended to the regime ${\rm Re}>1$ and large film thickness, provided that the dimensionless length scale $\tilde{\delta}$ is kept constant.

\section{Analysis of the energy dissipation}

To understand the dependence of the COR on various particle as well as liquid properties, it is helpful to analyze the associated energy dissipation. If $E_{\rm diss}$ is defined as the total kinetic energy loss of the particle during the impact, the dependence of the COR on the kinetic energy at impact $E_{\rm i}$ can be written as

\begin{equation}
\label{enEi}
e_{\rm n}=\sqrt{1-E_{\rm diss}/E_{\rm i}}.
\end{equation}

\noindent The dissipated energy $E_{\rm diss}$ can be treated as the sum of two parts: the part transferred into the solid body $\Delta E_{\rm dry}$, and the other part taken by the wetting liquid $\Delta E_{\rm wet}$, i.e.

\begin{equation}
\label{Esum}
E_{\rm diss}=\Delta E_{\rm dry}+\Delta E_{\rm wet}.
\end{equation}

\noindent Provided that the two parts are independent from each other, i.e.\ the wetting liquid does not change the energy dissipation of the dry impact, $\Delta E_{\rm wet}$ could be obtained experimentally by

\begin{equation}
\label{EwetExp}
\Delta E_{\rm wet}=E_{\rm i}(e_{\rm dry}^2-e_{\rm n}^2).
\end{equation}

The whole process of the colliding event can be separated into two parts: impact and rebound. During the impact, the kinetic energy of the particle will partly be transferred to the wetting liquid. This amount of energy will finally be dissipated by the motion of the viscous liquid, including surface waves or even splashes, depending on the competition between the inertial, viscous and surface forces. During the rebound, the rupture of the capillary bridge will lead to a certain amount of surface energy loss in addition to the damping caused by the motion of the liquid. Moreover, the mass of the wetting liquid dragged away by the sphere might lead to a further reduction of the COR. Based on the above analysis, one can take the most prominent terms and use

\begin{equation}
\label{Ewet3}
\Delta E_{\rm wet} \approx \Delta E_{\rm visc}+\Delta E_{\rm b}+\Delta E_{\rm acc}
\end{equation}

\noindent to estimate $\Delta E_{\rm wet}$ theoretically , where $\Delta E_{\rm visc}$ represents the energy dissipated via the viscous damping force acting on the particle, $\Delta E_{\rm b}$ corresponds to the energy loss arising from the surface energy change of the fluid, and $\Delta E_{\rm acc}$ is the kinetic energy change of the fluid before and after the colliding event.

In the limit that the thin film lubrication theory applies, the viscous force acting on the particle can be estimated by $F_{\rm v}=3\pi \eta D^2 v_{\rm impact}/2x$ \cite{Davis02}, where $x$ denotes the distance between the sphere and the plate. Following Ref. \cite{Ennis91}, one might assume the same force law for both approach and departure of the sphere, despite that the boundary condition for the latter case is dramatically different from the former one. By integrating over the distance that the viscous force applies, we obtain

\begin{equation}
\label{Visc}
\Delta E_{\rm visc}=\frac{3}{2}\pi \eta D^2 v_{\rm impact} (\ln{\frac{\delta}{\epsilon}}+\ln{\frac{\delta_{\rm r}}{\epsilon}}),
\end{equation}

\noindent where $\epsilon=5\,\mu m$ is the roughness of the sphere, and $\delta_{\rm r}$ is the rupture distance of the liquid bridge. For a crude estimation, we take a fixed $\delta_{\rm r}=2D$ according to the snapshots taken and assume that the velocity does not change during the impact.

\begin{figure}
\includegraphics[width = 0.4\textwidth]{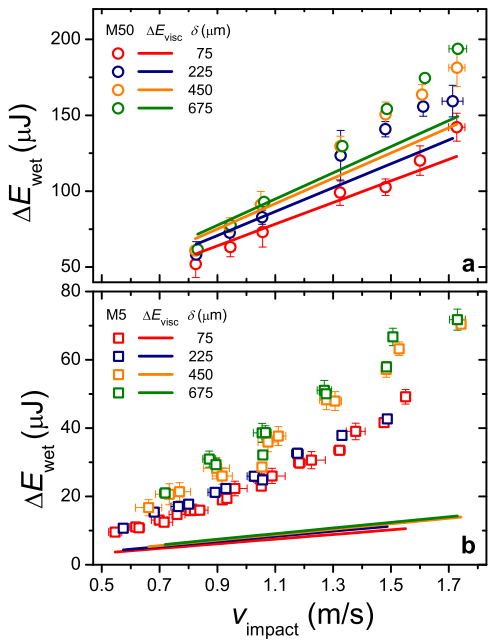}
\caption{\label{Ewet}(color online) Energy dissipation $\Delta E_{\rm wet}$ due to wetting  as a function of impact velocity with the wetting liquid properties the same as in Fig.\ref{vivo} for both silicone oil M50 (a) and M5 (b) wetting. Solid lines represent estimated values of the energy dissipation $\Delta E_{\rm visc}$ from viscosity (see text for detailed descriptions). The line colors are in accordance with the data points for various film thickness.}
\end{figure}

In Fig.\,\ref{Ewet}, $\Delta E_{\rm wet}$ for the experimental results shown in Fig.\,\ref{vivo} is plotted in comparison with $\Delta E_{\rm visc}$. Qualitatively, the monotonic growth of the energy dissipation with the impact velocity, and the increase of energy dissipation with the film thickness agree with the estimation from Eq.\,(\ref{Visc}). This growth with the impact velocity deviates slightly from a straight line, which is suggested by the model, indicating that the dominating energy dissipation term has a higher order dependence on the impact velocity. Quantitatively, a comparison between the estimated viscous damping term $\Delta E_{\rm visc}$ and $\Delta E_{\rm wet}$ reveals that a substantial amount of the latter can be attributed to the viscous damping for the case of silicone oil M50 wetting, while this term plays a much weaker role for the case of less viscous silicone oil M5 wetting. This could be understood in terms of the difference of the Reynolds number. For less viscous silicone oil M5 wetting, the range of Reynolds number is an order of magnitude larger than that for silicone oil M50 wetting. Thus the energy loss due to the inertia of the wetting liquid is more prominent. As a consequence, the estimated $\Delta E_{\rm visc}$ plays a less important role in the total energy dissipation $\Delta E_{\rm wet}$. Since there exists a systematic deviation of $\Delta E_{\rm wet}$ from the predicted $\Delta E_{\rm visc}$ with the growth of the impact velocity and the decrease of the viscosity, one could estimate the threshold Reynolds number below which the viscous effect dominates. Taking $|\Delta E_{\rm wet}-\Delta E_{\rm visc}|/\Delta E_{\rm wet}$ as the order parameters and $30\%$ deviation as the limit, one can estimate the corresponding Reynolds number to be ${\rm Re}\approx 10$ for the case of $\tilde{\delta}\approx 0.04$. 

The second term in Eq.\,(\ref{Ewet3}) stems from capillary forces. Upon rebound of the sphere, a liquid bridge may form between the sphere and the liquid surface. The corresponding energy dissipation due to the deformation and rupture of this liquid bridge can be estimated by an integration of the force arising from the surface tension over the length that it acts. This capillary force has two components: the surface tension acting on the perimeter of the neck ($2\pi r_{\rm n} \gamma$ with $r_{\rm n}$ the neck radius and $\gamma$ the surface tension), and the second part arising from the Laplace pressure $p_{\rm b}$ that acts on the cross-section of the neck ($-\Delta p_{\rm b} \pi r_{\rm n}^2$). Based on quasi-static experimental verifications, a close form approximation of the capillary force $F_c$ between two spheres has been given as

\begin{equation}
F_c=\frac{\pi D \gamma cos(\phi)}{1+2.1S^*+10S^{*2}},
\end{equation}

\noindent where $S^*=s\sqrt{D/2V_{\rm b}}$ is the half separating distance $s$ rescaled by the characteristic length scale $\sqrt{D/2V_{\rm b}}$ with the bridge volume $V_{\rm b}$, and $\phi$ corresponds to the contact angle \cite{Willet00}.

Taking the rupture distance $\delta_{\rm c}$ as the integration limit and assuming a contact angle of $0^{\circ}$, one could estimate the rupture energy of the liquid bridge to be

\begin{equation}
\Delta E_{\rm b}\approx \pi \gamma \sqrt{2V_{\rm b} D}.
\end{equation}

A rough estimation of the bridge volume $V_{\rm b}\approx D^3/16$, based on the snapshot taken, gives rise to $E_{\rm b}\approx 0.7 {\rm \mu J}$ for silicone oil wetting a glass bead with diameter $5.5\,{\rm mm}$. Considering the energy dissipation obtained by the COR measurements shown in Fig.\,\ref{Ewet}, $E_b$ plays a minor role for the few mm sized particle used here. Note that $E_{\rm b}$ plays a more prominent role as $D$ decreases, because its growth with $\sqrt{D}$ is in contrast to $E_{\rm visc} \propto D^2$.

As demonstrated in Fig.\ref{Ewet}, both the damping from the viscous force and the rupture of liquid bridges can not explain the amount of energy dissipation for the case of silicone oil M5 wetting. Therefore, other effects, like e.g. the inertia of the liquid or surface waves, should be considered.

As a first approximation, the inertial effect could be estimated from the kinetic energy of the liquid being pushed aside by the impact \cite{Vollmer10}. The volume of the liquid can be estimated by the spherical cap immersed in the liquid film $V=\pi D^3 \tilde{\delta}^2(1/2-\tilde{\delta}/3)$. From the length scale taken as the base radius of the spherical cap $\sqrt{1-(1-2\tilde{\delta})^2}D/2$ and the time scale $\delta/v_{\rm impact}$ for the particle to penetrate the liquid layer, one estimates the average velocity $v_{\rm l}=v_{\rm impact} \sqrt{1/\tilde{\delta}-1}$. As a consequence, the kinetic energy $\Delta E_{\rm acc}$ of the liquid being pushed aside yields

\begin{equation}
\label{Eacc}
\Delta E_{\rm acc}=\frac{1}{2}\rho_{\rm l}Vv_{\rm l}^{2}=3\tilde{\rho}(\tilde{\delta}-\frac{5}{3}\tilde{\delta}^2+\frac{2}{3}\tilde{\delta}^3) E_{\rm i},
\end{equation}

\vspace{1em}
\begin{figure}
\includegraphics[width = 0.4\textwidth]{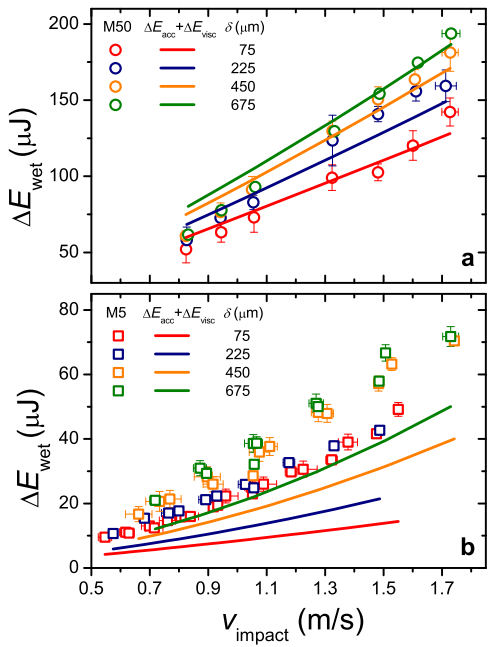}
\caption{\label{Ewet2}(color online) Data points are the same as shown in Fig.\,\ref{Ewet}. Solid lines represent estimated values with the consideration of both viscous damping $\Delta E_{\rm visc}$ and the energy transfer to the fluid $\Delta E_{\rm acc}$ (see text for detailed descriptions). The line colors are in accordance with the data points for various film thickness.}
\end{figure}

\noindent which shows a linear dependence on the kinetic energy $E_{\rm i}$ of the impact particle. Figure~\ref{Ewet2} shows that, by taking both $\Delta E_{\rm visc}$ and $\Delta E_{\rm acc}$ into account, the influence from the inertia effect is more prominent for less viscous silicone oil M5 wetting. The combination of both forces leads to a better agreement with the experimental data, when compared to Fig.\,\ref{Ewet}(b). However, considering both the inertial and the viscous damping parts of the energy dissipation can not explain the experimental results for less viscous silicone oil M5 wetting quantitatively. This indicates that further theoretical considerations, e.g. on additional energy dissipation terms, or a more careful characterization of the inertial effects, are desirable.

The fact that the ratio between $\Delta E_{\rm acc}$ and $E_{\rm i}$ is not velocity dependent suggests that the inertia of the wetting liquid will not contribute to the impact velocity dependence of the $e_{\rm n}$. It does, however, explain why $e_{\rm inf}$ obtained from linear fits of the data is generally smaller than $e_{\rm dry}$. Based on the Eqs.\,(\ref{EwetExp}) and (\ref{Ewet3}), the $E_{\rm i}$ dependent COR could be written as

\begin{equation}
e_{\rm n}=\sqrt{e_{\rm dry}^2-\frac{\Delta E_{\rm acc}}{E_{\rm i}}-\frac{\Delta E_{\rm visc}}{E_{\rm i}}-\frac{\Delta E_{\rm b}}{E_{\rm i}}}.
\label{en}
\end{equation}

In the limit of large $v_{\rm impact}$, the last term $\Delta E_{\rm b}/E_{\rm i}$ could be safely ignored so that two independent parameters are enough to determine the impact velocity dependence of the COR. A comparison to the linear fits $e_{\rm n}=e_{\rm inf}(v_{\rm impact}-v_{\rm c})$ employed before immediately reveals that the linear fit is a first order approximation of the Eq.\,(\ref{en}) and $e_{\rm inf}=\sqrt{e_{\rm dry}^2-\Delta E_{\rm acc}/E_{\rm i}}$. By ignoring the higher order terms of $\tilde{\delta}$ in $\Delta E_{\rm acc}$, one derives a linearized form $e_{\rm inf}=e_{\rm dry}-3\tilde{\rho}\tilde{\delta}/2e_{\rm dry}$, which suggests $e_{\rm inf}\approx0.92$ for typical experimental values of $\tilde{\rho}= 1/2.5$ and $\tilde{\delta}= 0.1$. In comparison to Fig.\,\ref{LinFit2}(b), this estimated value is close to the $e_{\rm inf}$ obtained from fitting. Moreover, the monotonic decrease with $\tilde{\delta}$ is captured by this formula qualitatively, except for the $75\,\mu$m thick silicone oil M5 wetting case.

\section{Conclusion}

In summary, the normal coefficient of restitution (COR) for a free falling sphere on a wet surface is investigated experimentally. The dependence of the COR on the impact velocity and various particle and liquid film properties is discussed in relation to the energy dissipation associated with the impact process.

i) For dry impact, the COR corresponds to the slope of the rebound vs. impact velocity. For wet impacts, the rebound velocity and the impact velocity are also found to fall into a straight line, but with a smaller slope and an offset corresponding to a finite critical impact velocity. Even though linear fitting is only a first order approximation of $e_{\rm n}$, it successfully characterizes the impact velocity dependence of the COR with two parameters $e_{\rm inf}$ and $E_{\rm c}$. Therefore this simplification is justified to be a good candidate for computer simulations aiming at modeling wet granular dynamics on a large scale.

ii) The dependence of the COR on the impact velocity, dimension of the sphere and the viscosity of the liquid could be well characterized by the Stokes number, which is defined as the ratio between the inertia of the sphere and the viscosity of the liquid, provided that the dimensionless length scale $\tilde{\delta}$ is fixed. This result supports the usage of the Stokes number for scaling the data, even beyond the low Reynolds number regime where it has originally been introduced.

iii) Concerning the energy dissipation arising from the wetting liquid, the viscous damping term dominates for Reynolds number up to ${\rm Re}\approx 10$. Away from that limit, further effects, such as the inertia of the liquid film, have to be considered. The rupture energy of a capillary bridge during the rebound process could be safely ignored for the few mm sized particles used here.

The above conclusion suggests that further investigation on the dynamics of wet impacts is desirable for a better understanding of the COR and the energy dissipation associated. This requires an accurate determination of the particle trajectories during the impact with the liquid film experimentally, as well as a comparison with numerical simulations (see e.g. \cite{Bauer00}).

The authors would like to acknowledge Mario Sch\"orner for the help in the construction and calibration of the film thickness measurement part of the setup, and Bryan J. Ennis and J\"urgen Vollmer for helpful hints. This work is partly supported by the German Science Foundation within Forschergruppe 608 `Nichtlineare Dynamik komplexer Kontinua' through Grant No. Kr1877/3-1.

%

\end{document}